\documentclass{elsart}
\usepackage{amssymb}
\usepackage{graphicx}
\usepackage{dcolumn}
\usepackage{bm}
 
\begin{document}
  
\begin{frontmatter}
 
\title{Ab initio study of the double row model 
of the Si(553)-Au reconstruction}
   
\author{Sampsa Riikonen}
\ead{swbriris@sc.ehu.es}
\address{Universidad del Pa\'{\i}s Vasco, 
Facultad de Qu\'{\i}micas, Departamento de F\'{\i}sica de Materiales, 
Apdo 1072,
20080 San Sebasti\'an, Spain and 
Donostia International Physics Center (DIPC),
Paseo Manuel de Lardizabal 4, 20018 San Sebasti\'an, Spain}
\author{Daniel S\'anchez-Portal}
\ead{sqbsapod@sc.ehu.es}
\address{Unidad de F\'{\i}sica de Materiales, 
Centro Mixto CSIC-UPV/EHU, 
Apdo 1072, 20080 San Sebasti\'an, Spain
and
Donostia International Physics Center (DIPC),
Paseo Manuel de Lardizabal 4, 20018 San Sebasti\'an, Spain}

\begin{abstract}
Using x-ray diffraction Ghose {\it et al.}
[Surf. Sci. {\bf 581} (2005) 199]
have recently 
produced a structural model 
for the quantum-wire surface Si(553)-Au. 
This model presents two parallel gold wires located
at the step edge. Thus, the structure and the  
gold coverage are quite different
from previous proposals. 
We present here an \emph{ab initio} study using density
functional theory of the 
stability, electronic band structure and 
scanning tunneling microscopy images of this
model.
\end{abstract}
 
 
\end{frontmatter}
 
\section{Introduction}

One of the most interesting properties of the one-dimensional
metals is the breakdown of the Fermi-liquid behavior.
The nature of the low-energy excitations is in this case
quite different from the expectations based on the Fermi liquid
theory. The single-particle excitations have to be
replaced by separate spin and charge collective
excitations~\cite{Giamarchi04}. This collective
behavior leads to several exotic phases at low temperatures.
However, although these theoretical
predictions are clear and well-founded, the observation
of the spin-charge separation and other effects has proven quite elusive.
One of the main problems to observe this
metallic and non-Fermi-liquid ground state (the so-called Luttinger
liquid) is the
instability of the one-dimensional metals against
the Peierls distortion~\cite{peierls}
and, in some cases, the Mott insulating state~\cite{Mott}.
 
Atomic-thin gold wires are spontaneously formed
on several vicinal Si(111) surfaces after the deposition of
gold in the sub-monolayer regime.
Each terrace contains at least one of these gold wires 
running parallel to the step-edge. 
These surface reconstructions have been proposed as 
excellent model systems to probe the electronic
properties of one-dimensional metals
and, consequently, 
have attracted much attention 
in recent years~\cite{himpsel_review}.
Some examples that have been extensively studied
are the 
Si(557)-Au~\cite{557_first,luttinger4,557_bands,daniel1,557_xray,daniel2,557_wires,sanchez_riikonen,Yeom05}
and the Si(111)-(5$\times$2)-Au~\cite{stm_5x2_2,marks_plass,1d_states,5x2_bands,5x2_corr,5x2_comp,doping,arpes_latest,adatom_wires,5x4_stab,our_5x2}
reconstructions.
These surfaces have been suggested to present several 
advantages over
other quasi-one-dimensional systems: {\it i})
the gold wires 
self-assemble under the appropriate conditions thus, once
these conditions have been determined, clean and well-ordered
samples can be easily fabricated; {\it ii})
vicinal surfaces act as templates creating regular
arrays of wires which are suitable for high resolution
studies using angle-resolved photoemission;
{\it iii}) furtheremore,  
inter-chain distances and interactions 
can be tunned by changing the 
average terrace width, which can be controlled
by the miscut angle; {\it iv}) due to the interaction
of the gold atoms with the substrate these
structures are expected to be more robust against
Peierls-like distortions (remaining metallic) as compared to other
quasi-one-dimensional systems like 
e.g. organic compounds; {\it v}) 
the substrate is a semiconductor so the 
bands appearing near the Fermi energy associated with 
the gold chains should reflect, at least
to some extent, the properties of
one-dimensional metals.
Unfortunately some of these expectations have not been
fully confirmed by the experiment. For example, the 
Si(557)-Au has been reported to suffer a 
Peierls-like transition to a semiconducting 
ground state~\cite{557_wires}, and the 
semiconducting or metallic character of the
Si(111)-(5$\times$2)-Au surface is still a matter
of debate~\cite{5x2_bands,arpes_latest,5x4_stab}.

In this context, the band structure
of the Si(553)-Au surface was considered
a very interesting opportunity~\cite{himpsel_review,553}. 
This surface presents three dispersive surface bands
with clear one-dimensional character.
Two of them are almost identical to 
the nearly half-filled bands found 
for the Si(557)-Au surface~\cite{557_bands}.
Therefore, in analogy with this surface, these bands 
can be assumed to appear as a consequence of the spin-orbit splitting
of a dispersive surface band associated with 
the gold wires~\cite{sanchez_riikonen}. 
The third band is metallic at room temperature and nearly 
1/4 filled~\cite{553}, its origin being unclear.
This quarter-filled band could create an unique opportunity
for observing the Luttinger liquid behavior. 
This is in contrast with half-filled
bands, which are unstable against a Mott transition for
large values of the electron-electron interaction,
preventing the possible
observation of a Luttinger metal~\cite{himpsel_review}.
Furthermore, having several surface bands with different
fractional fillings could prevent
the appearance of a Peierls instability if all of these
bands come from the same wire in the surface.
Again some of these expectations
seem to have been frustrated by the experimental evidence.
A very recent study indicates that both groups of 
bands (the half-filled ones and the nearly 1/4 filled bands) 
come from different regions of the surface unit cell 
and suffer independent Peierls-like instabilities 
with different transition temperatures~\cite{Ahn553}. 

The origin of the 
quarter-filled band in the Si(553)-Au surface
is an interesting line of 
research. 
If this band 
were understood, then it might 
be possible to envision ways to modify the surface or 
to fabricate analogous 
structures on different substrates such that
one could finally observe the
Luttinger metal behavior. One important step in such
research program is to find
a reliable structural
model. This is a necessary prerequisite to understand the electronic
structure of any complex surface reconstruction.
To date the atomic structure 
of the Si(553)-Au reconstruction
has not been completely established. 
Until quite recently all the available structural information
was based on scanning tunneling microscopy (STM) 
studies~\cite{553,Ahn553,Crain05} that, however, were
not very conclusive. The reported gold coverage 
was $\sim$0.24~ML~\cite{553}, which pointed to the
existance of a monatomic gold wire in each terrace on the surface. 
According to this information, and to a plausible 
analogy with other systems
like the Si(557)-Au surface, a series of structural models
were investigated by Riikonen and S\'anchez-Portal~\cite{riikonen_553}
and Crain {\it et al.}~\cite{himpsel_review}. However, 
none of these models provided a 
fully satisfactory description
of the observed band 
structure and STM images~\cite{riikonen_553}. 
More recently,
Ghose {\it et al.}~\cite{553_robinson} have 
proposed a detailed structural model of 
the Si(553)-Au surface reconstruction
based on an x-ray diffraction study. Surprisingly, their model
contains twice as much gold ($\sim$0.54~ML) as the original proposals.
According to these authors, the gold coverage is firmly
established by their analysis of the x-ray data~\cite{553_robinson}. 
However, it seems
quite unlikely that the gold dose were so severely underestimated
in the previous studies~\cite{553}. Therefore, this might imply 
that there exist two stable structures, with different gold content,
for the Si(553)-Au surface.

In this work, we study the geometry and the electronic properties of
the model proposed by Ghose {\it et al.}~\cite{553_robinson} 
using {\it ab initio} density functional calculations.  
We perform constrained
and unconstrained relaxations, starting from the structural 
coordinates of Ghose and collaborators~\cite{structure}.
We analyze the stability, electronic band structure and 
simulated STM images of the model.  In the light of our calculations,
this geometry is not stable and the predicted
band structure is quite different from that reported
in the photoemission experiments.
 
\section{Computational method}
 
All the calculations were performed using
the SIESTA code~\cite{siesta1,siesta2,reviewsiesta}
We used the local density approximation~\cite{lda} and
norm conserving pseudopotentials~\cite{tm}. The gold
pseudopotential included scalar relativistic effects
and was similar to that used in 
Ref.~\cite{au1} and 
in our previous calculations~\cite{our_5x2,riikonen_553}.
A double-$\zeta$ (DZ)
basis set of atomic orbitals (i.e. including two
different radial functions for the
3$s$ orbitals, and another two to represent
the 3$p$ shell) was used for silicon.
The gold basis included doubled and polarized
6$s$ orbitals (thus including a single
6$p$ shell) and a single 5$d$ shell. 
Although a DZ basis is usually sufficient to obtain a quite
good description of the occupied electronic states and
the relaxed geometries in silicon systems, the use of
a more complete basis set is necessary to describe the
unoccupied part of the band structure even at low energies.
For this reason we use a a double-$\zeta$
polarized (DZP) silicon basis set for the calculation
of the band structure and the STM images.
We used an energy shift of 200~meV to define the
cut-off radii of the different orbitals~\cite{siesta2}.
The corresponding radii are
5.25 and 6.43 Bohr for the 3$s$ and 3$p$~(3$d$) Si
orbitals, and 6.24 and 4.51 Bohr for the 6$s$~(6$p$) and 5$d$ Au
orbitals, respectively.
                                                                                
We modelled the surface using a slab formed by
three silicon bilayers. The 
bottom silicon layer is saturated with hydrogen.
To avoid artificial stresses the
lateral lattice parameter was fixed to the bulk theoretical
value calculated with similar approximations (5.48~\AA\ with 
a DZ basis set).
The structures were relaxed until the maximum force
component was less than 0.04 eV/\AA.
The distance between neighbouring slabs was 15~\AA. 
A 4$\times$4 sampling of the surface Brillouin
zone and a real-space grid equivalent to a
100~Ry plane-wave cut-off was used.
The Tersoff-Hamann theory was used to produce the simulated
STM images \cite{stm_theory}.
 
\section{Results}
 
\begin{figure*}
\begin{center}
\includegraphics[keepaspectratio,width=5cm]{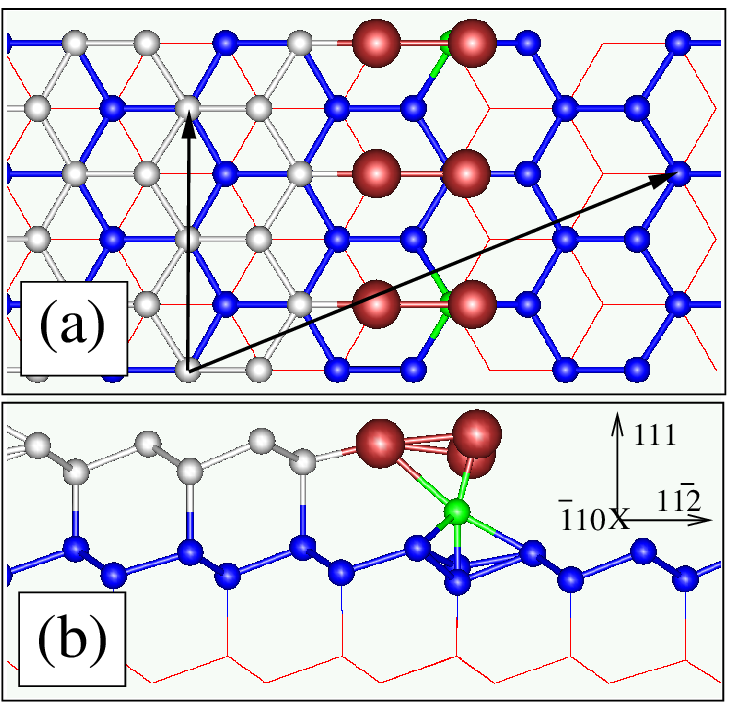}
\includegraphics[keepaspectratio,width=5cm]{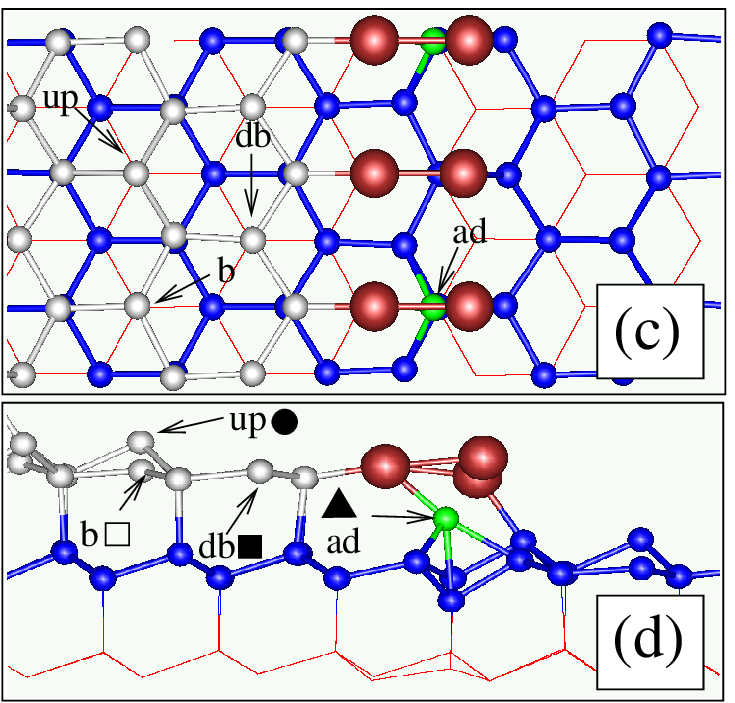}
\includegraphics[keepaspectratio,width=10cm]{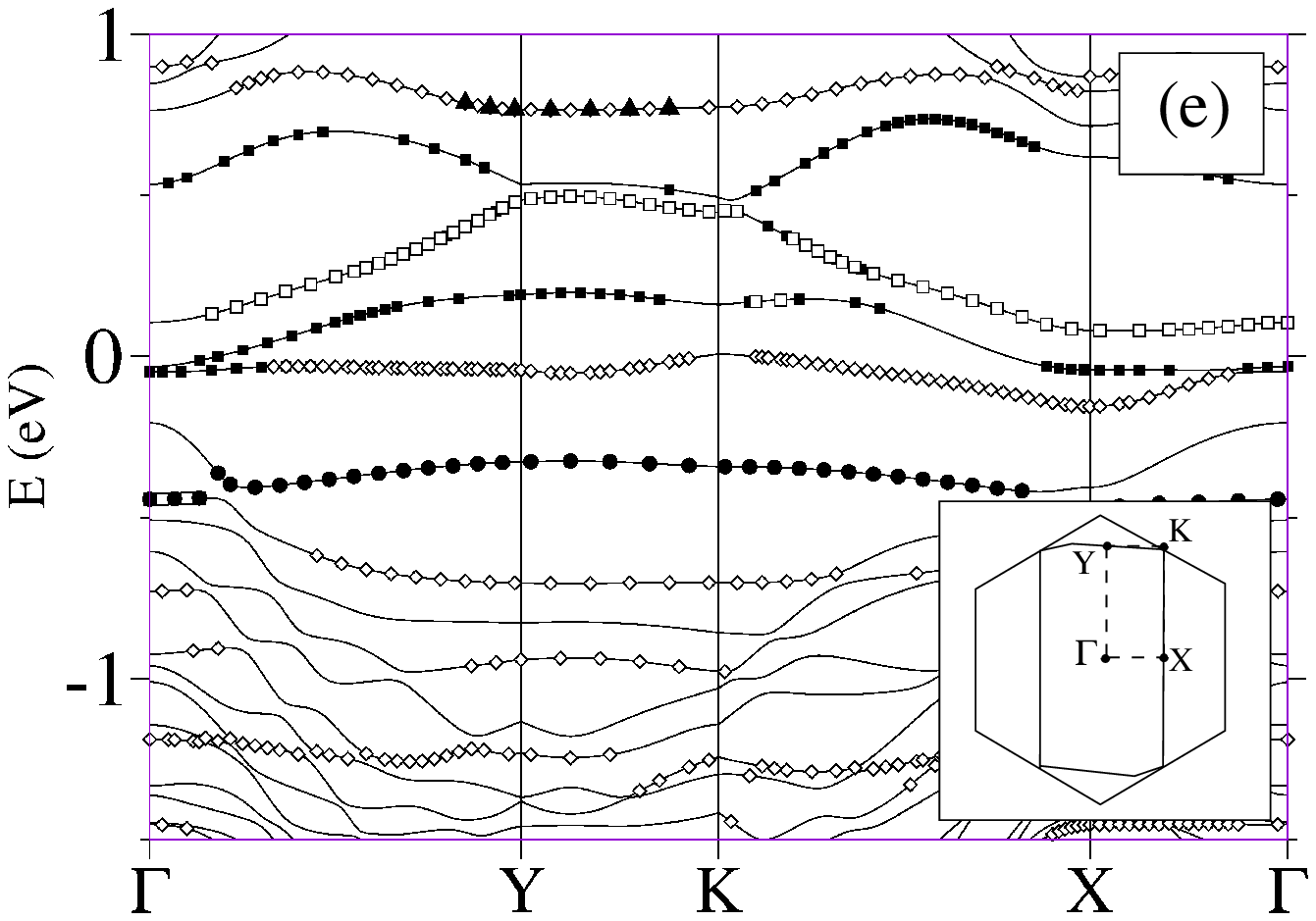}
\end{center}
\caption{\label{fig:constr}
The model of Ghose {\it et. al.}~\cite{553_robinson} for the Si(553)-Au
reconstruction (a and b, panel (a) also shows the unit cell vectors). 
The same structure after constrained relaxation (c and d) 
and the corresponding band structure (see text).  The main atomic 
character of the surface bands is
indicated with different symbols in panel (e) which correspond 
to those used to label different atoms in panels (c) and (d).
The diamonds correspond 
to the gold atoms in the step edge and their neighboring silicon atoms.
The inset of panel (e) illustrates the Brillouin-zone of the Si(553)-Au 
reconstruction. The Brillouin-zone of a 2$\times$2 supercell of
the unreconstructed 
Si(111) surface is also shown for comparison.
The $\Gamma$-Y and K-X directions
are parallel to the gold wires.}
\end{figure*}

The proposal for the structure of the Si(553)-Au reconstruction 
by Ghose {\it et. al.}~\cite{553_robinson}
can be seen in Fig.\ref{fig:constr} (a) and (b). The main features 
are the double row of gold 
atoms located at the step edge of the Si(553) surface
and the silicon adatoms residing right below some of these gold atoms. 
This reconstruction is
quite different from the other and better known structures induced
by the deposition of gold 
on vicinal Si(111) surfaces like, for example, the Si(557)-Au 
reconstruction~\cite{daniel2,himpsel_review}. 
Particularly surprising is the position of the gold atoms 
at the step edge. It has been shown by density functional
calculations in several 
similar surfaces that the silicon subsitutional sites in the
middle of the terraces are typically
more favorable for gold~\cite{daniel1,daniel2,riikonen_553}.  
Another striking fact is the very large distance between the gold atoms
along the step edge (see Fig.~\ref{fig:constr}~(a)). This distance
($\sim$3.8~\AA) has to be compared, for example, to the nearest neighbor
distance in bulk gold (2.9~\AA). 
In the direction perpendicular
to the step edge we find two slightly different Au-Au distances,
$\sim$2.7~\AA\ and $\sim$2.8~\AA. These distances are intermediate between the 
bond length of the gold dimer (2.5~\AA) and that of bulk. 
Another peculiarity of this structure is that the silicon 
terrace remains basically unreconstructed. This is in clear constrast
with other systems like the Si(557)-Au 
surface~\cite{daniel1,557_xray,daniel2}, the 
Si(111)-(5$\times$2)-Au~\cite{5x2_comp,doping,our_5x2}, and even previous
theoretical
models of the Si(553)-Au~\cite{riikonen_553}. For example, the so-called
honeycomb chain reconstruction of silicon~\cite{honeycomb}
is known to occur in most of the gold induced reconstructions
on vicinal Si(111)~\cite{himpsel_review}. However, it is absent 
in the model studied here. Thus, the gold double-row 
model proposed by Ghose {\it et. al.} can be pictured 
as a collection of gold dimers
attached to the edges of the terraces of a largely unreconstructured
Si(553) surface.
The gold dimers are oriented along the normal to the step edge. 
There are two types of gold dimers. This configuration 
can be justified for one of these dimers, which bonds to a 
silicon adatom in the terrace below with a reasonable Si-Au distance
of $\sim$2.4~\AA. However, this arrangement
seems rather artificial and unstable for the other dimer. 
We performed structural relaxations 
to study the stability of this structural model. As we will see  
below the model turns out to be unstable and its
structure is greatly modified during the relaxation.
One could always argue that this result is a
pathology of the local density approximation or other
approximations used in this work.
For this reason we have perfomed constrained relaxations
that, while optimizing some of the bond lengths and 
bond angles, preserve the main characteristics of the 
structure in Ref.~\cite{553_robinson}. The electronic 
band structure and the simulated STM images are then 
calculated for this optimized structure and compared
to the available experimental information.

Fig.~\ref{fig:constr} (c) and (d) shows the result
of a constrained relaxation in which the relative positions
of the gold atoms are not allowed to change (i.e. the
gold atoms cannot move respect to each other).
All other degrees of freedom are optimized: {\it i}) the position 
of the center of mass of the gold atoms and, {\it ii}) the positions of all the
silicon atoms in the slab, except those in the lowest
layer which remain in perfect bulk positions. 
As a stronger scatterer, the gold
positions should be the most reliable in the 
experiment~\cite{557_xray,553_robinson}. This justifies 
the approach followed here.
After this constrained relaxation, 
the silicon atoms of the first layer reconstruct
to some extent. The atoms labeled 
``up'' and ``b'' (see Fig.~\ref{fig:constr}
(c) and (d)) give rise to a buckling of the surface, a
well known silicon reconstruction~\cite{buckle} in which 
there is a charge transfer from the lower atom to the the elevated one.
This is clearly reflected in the
electronic band structure 
shown in Fig.~\ref{fig:constr}~(e). The ``up'' atom creates
a fully occupied band with small dispersion (solid circles), while 
a more dispersive unoccupied band (open squares) is 
associated with the ``b'' atom.
Atom labeled ``db''  has a partially occupied
dangling-bond. The corresponding dispersive metallic band 
(solid squares) can be found close
to the Fermi level in  Fig.~\ref{fig:constr}~(e). 
Several surface bands appear associated with the 
gold atoms and their neighboring silicon atoms in the 
step edge (open diamonds). 
However, all these bands are quite flat.
This is in contrast with the band structures of other 
reconstructions of gold in vicinal Si(111). In those cases
the gold atoms occupy silicon substitutional positions in the middle
of the terraces and produce quite dispersive one-dimensional bands
that dominate the photoemission 
spectra~\cite{daniel2,sanchez_riikonen,doping,our_5x2,riikonen_553}.
Furthermore, in Ref.~\cite{sanchez_riikonen} it was shown that the
presence of gold induces a spin-orbit splitting of the hybrid silicon-gold
bands that explains
the observation of two proximal one-dimensional
bands in the Si(557)-Au surface~\cite{557_bands,557_wires}. 
The photoemission of the Si(553)-Au surface also 
shows two proximal half-filled
bands similar to those of the Si(557)-Au~\cite{himpsel_review,553,Ahn553}.
Therefore, it is tempting to associate these 
bands with the gold wires and their
silicon neighbors in analogy to the case of Si(557)-Au. Since
in the present calculations we are not including the spin-orbit 
interaction, these 
two proximal bands should appear as a single
dispersive band~\cite{sanchez_riikonen}. 
Unfortunately, a dispersive
band associated with the gold atoms 
is completely absent in Fig.~\ref{fig:constr}~(e).
The band coming from the partially occupied dangling-bonds in the 
``db'' atoms could be identified with the $\sim$1/4 filled band
of the Si(553)-Au~\cite{himpsel_review,553,Ahn553}. However, 
this identification is also not very clear since in the experiment
this band goes down to much lower energies.
We can thus conclude that the band structure calculated for the
model proposed by Ghose {\it et al.} fails to reproduce the photoemission data.
Of course, given the discrepancy in the gold coverage reported
in the photoemission work~\cite{himpsel_review,553} and the x-ray 
diffraction work of Ghose {\it et al.}~\cite{553_robinson},
it is perfectly plausible that 
we are dealing with different reconstructions of the surface.
In such case, the data reported in Fig.~\ref{fig:constr}~(e)
can be considered as the predicted electronic band structure
for the double row model proposed by Ghose {\it et al.}
using the local density approximation. 

\begin{figure}
\begin{center}
\includegraphics[keepaspectratio,width=5cm]{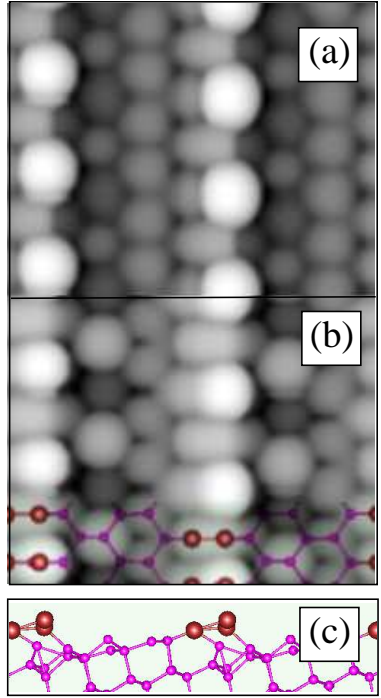}
\end{center}
\caption{
\label{fig:stm}
Simulated STM images of the double row model~\cite{553_robinson} 
of the Si(553)-Au reconstruction after constrained relaxation: 
panel (a) for a +1.0~V bias voltage (empty states),
and (b) for -1.0~V bias voltage (filled states). Panel (c)
shows the corresponding atomic configuration viewed from the side.}
\end{figure}

Simulated STM images for filled and empty states are presented in
Fig.~\ref{fig:stm}.  The gold atoms show as alternating bright spots 
along the [$\bar{1}10$] direction with a $\times$2 periodicity.  
This periodicity 
reflects the alternating heights of the gold atoms induced
by the presence of a row of silicon adatoms below them.  
Another feature with $\times$2
periodicity is seen in the middle of the terrace
as a result of the buckling of the silicon surface layer.
In spite of the difference in the reported gold coverages
we can insist in comparing with the available 
experimental images~\cite{himpsel_review,553,Ahn553,Crain05}.
At room temperature the step edge is observed in the experiment
as a continous bright line.
Another less pronounced feature is found in the middle of the terrace with 
a $\times$2 modulation already at room temperature. At low temperature
the terrace chain shows a more clear  $\times$2  periodicity, while the
line at the step edge develops a $\times$3 modulation. While
the doubling of the periodicity in the middle of the terrace
is reproduced by the model studied here, the image produced by the 
step edge is quite different. The appearance of bright spots
in the step edge is linked to the presence of the silicon adatom
in the terrace below.
One could then speculate on creating a better agreement with the STM
images by introducing an adatom only every three unit cells. However, 
this could hardly produce the observed temperature variation.
We can thus conclude that the STM images predicted for 
the double row model of the Si(553)-Au reconstruction differ
considerably from the reported STM images.

So far we have analyzed the results obtained for a structure
optimized under the restriction that the gold atoms remain
at the experimentally determined positions. We can now release this
constraint and, starting from this partially relaxed structure,
fully optimized the geometry of the surface. 
By doing this we discover that the proposal of ŋGhose {\it et al.}
is not stable, at least within our computational approach.
Although we do not find strong changes in the silicon terrace,
the structure of the gold double-row is completely modified.
This is clearly seen in Fig.~\ref{fig:last}, where we
show the structure
of the surface after 300 steps of unconstrained structural 
relaxation. The gold atom that was initially sited on top 
of the silicon adatom has moved to a new position on top of 
the neighboring rest-atom. The configuration of 
the silicon adatom has also changed
considerably. The adatom moves to a higher position, its
height over the terrace being now comparable to that 
of the gold atoms.
This movement is possible because the adatom 
breaks a bond with one of the silicon surface atoms
and adopts a bridge-like configuration.
This broken bond is replaced by a new Si-Au bond. 
Although the structure shown in Fig.~\ref{fig:last}
is not completely 
relaxed, it becomes clear that the model 
of the surface proposed
in Ref.~\cite{553_robinson}, 
based on a
silicon step edge decorated with gold dimers,
is not stable. In particular,
the adsorption of one of the gold atoms on 
top of a silicon adatom is avoided. This is consistent
with previous density functional 
calculations~\cite{daniel2,doping}. In these calculations
it was shown that the adsorption of gold as an adatom
over the silicon surface is 
quite unfavorable compared to the substitution
of the gold atoms in the surface layer.

\begin{figure}
\begin{center}
{\it Fig.3 was attached to the submission as a jpeg file.
Sorry for the inconvenience.}
\end{center}
\caption{ Top (a) and lateral (b) view of the double row
model~\cite{553_robinson} of the Si(553)-Au after
300 steps of unconstrained structural relaxation.
\label{fig:last}}
\end{figure}
 
\section{Conclusions}
In this work we present \emph{ab initio} density 
functional calculations 
of the double row 
model proposed by Ghose {\it et al.}~\cite{553_robinson}
for the Si(553)-Au reconstruction. We address the stability 
of the model, as well as its electronic band structure
and STM images. Using the geometry obtained in a constrained structural
relaxation, which preserves the main characteristic 
of the proposal of Ref.~\cite{553_robinson}, we calculate
the band structure and STM images. 
We only find a dispersive
band with fractional filling close to the
Fermi level. This band comes from the silicon
dangling-bonds in the surface and its energy position 
and filling seems quite different from the bands
observed in the photoemission experiments~\cite{553,Ahn553}.
Dispersive bands associated with the gold atoms and
their silicon neighbors are completely absent,
which also seems to be in disagreement with the 
experimental evidence~\cite{sanchez_riikonen,553,Ahn553}.
At variance with the room temperature 
experimental STM images~\cite{553,Ahn553}, 
our 
simulated STM images do not show the step edge as a continuous
bright line, but exhibit a $\times$2 modulation associated
with the presence of the adatoms in the neighboring 
terrace. In the low
temperature experimental images
the step edge develops a $\times$3 periodicity~\cite{Ahn553}.
It 
might be possible to induce this $\times$3 periodicity in our
calculated STM images
by modifying the adatom content. However, it is not clear
how this could reproduce
the temperature dependence. In summary, the calculated
band structure and STM images for the model
proposed in Ref.~\cite{553_robinson} do not provide
a good agreement with the available experimental information
for this surface. Of course, 
it might be argued that this is a consequence
of the different gold coverage in the different
experimental approaches~\cite{553,553_robinson}. In fact, 
it is possible that the surface reconstructions 
studied by x-ray diffraction
in Ref.~\cite{553_robinson} and by photoemission and 
STM in references~\cite{553},~\cite{Ahn553}, and~\cite{Crain05}
are different. Unfortunately, the structure provided
by Ghose {\it et al.}~\cite{553_robinson} is unstable, at least
at the level of the local density approximation. 
When the geometry is relaxed without any constraints the structure
of the gold double-row attached to the step edge severely 
modifies from the proposal of Ref.~\cite{553_robinson}.
Therefore, we propose that the data of Ghose {\it et al.}
should be reanalyzed in the light of the present results 
and new proposal for the structure of the Si(553)-Au surface
obtained.

\section*{Acknowledgments}
The authors acknowledge illuminating discussions
with I. K. Robinson, F. J. Himpsel and J. E. Ortega.
This work was supported by the Basque Departamento de Educaci\'on,
the UPV/EHU (Grant No. 9/UPV 00206.215-13639/2001),
the Spanish Ministerio de Educac\'on y Ciencia
(Grant No. FIS2004-06490-C3-02), and the
European Network of Excellence FP6-NoE ``NANOQUANTA" (500198-2).
SR also acknowledges support from the Magnus Ehrnroot Foundation.
                                                                                 
\bibliographystyle{elsart-num}

\end{document}